\documentclass{article}
\usepackage{tikz}
\usetikzlibrary{positioning, calc}
\usepackage{amsfonts}
\usepackage{amsmath}
\usepackage{amssymb}
\usepackage{color}
\usepackage{eurosym}
\usepackage{hyperref}

\newtheorem{theorem}{Theorem}

\newtheorem{proposition}{Proposition}

\newtheorem{example}{Example}

\begin{document}

\title{The economics of sportscast revenue sharing\thanks{%
Financial support from grants PID2020-113440GBI00 funded by MCIN/AEI/
10.13039/501100011033; PID2023-146364NB-I00 funded by
MICIU/AEI/10.13039/501100011033, FEDER, and UE; and Contract-Program funded
by Universidade de Vigo is gratefully acknowledged. }}
\author{\textbf{Gustavo Berganti\~{n}os}\thanks{%
ECOBAS, Universidade de Vigo, ECOSOT, 36310 Vigo, Spain.} \\
\textbf{Juan D. Moreno-Ternero}\thanks{%
Department of Economics, Universidad Pablo de Olavide, 41013 Seville, Spain}}
\maketitle

\begin{abstract}
Sports are one of the most significant products of the entertainment
industry, accounting for a large portion of all television (and even
platform) viewing. Consequently, the sale of broadcasting and media rights
is the most important source of revenue for professional sports clubs. We
survey the economic literature dealing with this issue, with a special
emphasis on the crucial problem that arises with the allocation of revenues
when they are raised from the collective sale of broadcasting rights.
\end{abstract}

\medskip{} 

\medskip {}

\newpage

\section{Introduction}

Several professional sports leagues worldwide 
have at least 1 billion viewers. 
These viewers constitute a source of massive revenues, crucial for the
management of sports organizations. 
Paradoxically, it has been argued that free-to-air sports broadcasting
provided the foundations on which the highly commercialized sports industry
of the 21st century is built (e.g., Smith et al., 2015). Now, although a rising concern for the sports
industry is the behavior of young consumers, who have grown up in a digital
world with abundant access to free content, broadcasting contracts still
offer sizable amounts.
For instance, the current 2025-2026 season is witnessing the beginning of an
11-year agreement to air all of the NBA's nationally televised games that
will net the NBA roughly \$76 billion.\footnote{\url{https://www.espn.com/nba/story/_/id/40635523/faq-nba-signed-new-deal-disney-nbc-amazon-prime}}
The NFL also secured recently its media rights agreements 
collectively worth almost \$110 billion over 11 years, more than doubling
the value of the league's previous contracts, which expired in 2022.\footnote{\url{ 
https://www.spglobal.com/market-intelligence/en/news-insights/research/as-nfl-revenue-rises-current-media-rights-deals-ensure-future-success}}
Almost simultaneously, the English Premier League confirmed it would extend
its television deal, obtaining more than $\pounds 10$ billion for each of
the three upcoming seasons.\footnote{\url{https://theathletic.com/4240951/2023/03/08/premier-league-tv-rights-how-work-cost/}}

It has long been argued that broadcasting rights are the largest component
of revenues for major sports in rich nations. 
It is also well known that bundling products may increase revenue with
respect to selling products independently (e.g., Adams and Yellen, 1976) and
instances in which bundling occurs abound in real life (e.g., Ginsburgh and Zang, 2003; Berganti\~{n}os and Moreno-Ternero, 2015). 
This might partially explain the fact that the sale of broadcasting rights
for sports leagues is often carried out through some sort of collective
bargaining. Szymanski (2003) discusses the different policies of selling broadcasting rights in Europe and North America. Falconieri et al., (2004) provide a welfare analysis of
collective vs. individual sale of TV rights. Noll (2007) argues that
consumers are better off if television is competitive and leagues do not
centralize rights sales, concluding that centralization of rights sales does
not improve competitive balance or benefit financially weak clubs. Peeters
(2012) studies how media revenue sharing acts as a coordination device in
sports leagues. 

The ensuing sharing process among participating clubs is a
complex problem 
and sharing rules vary across the world. North America's one-team-one-vote environment paves the
way for equal sharing as the national contract can be approved only if there
is a virtual consensus among league clubs (e.g., Fort and Quirk, 1995). In contrast, 
performance-based reward schemes are widespread in Europe, which is rationalized by
the fact that European leagues compete for talent (e.g., Palomino and
Szakovics, 2004). Now, even within European leagues rules vary considerably. For instance,
Barcelona and Real Madrid, the two Spanish giant football clubs, used to
earn each more than $20\%$ of the revenues generated by the Spanish Football
League (e.g., Berganti\~{n}os and Moreno-Ternero, 2020a). In England,
however, the two clubs earning the most only made together $13\%$ of the
revenues generated by the English Premier League. This might partly explain
why in the last 21 editions of the Spanish Football League only twice the
champion was neither Barcelona nor Real Madrid (it actually was Atl\'{e}tico
de Madrid), whereas the English Premier League witnessed 4 different
champions (Manchester United, Manchester City, Leicester and Chelsea) in the
4 editions from 2013 to 2016 (and Liverpool won in 2020).
The impact of a revenue sharing arrangement on the competitive balance in a sports league has long been analyzed, with conflicting conclusions (mostly due to endorsing different approaches, models, or methodologies). 
Fort and Quirk (1995) famously concluded, under the assumptions of purely profit-maximizing clubs and Walrasian behavior, that gate revenue sharing (a specific form of revenue sharing) has no effect on competitive balance. 
K\'{e}senne (2000) argued that such conclusion did not hold in a league with purely win-maximizing clubs. Szymanski and K\'{e}senne (2004) defended that, under the assumptions of purely profit-maximizing clubs and Nash behavior, increasing gate revenue sharing among clubs in a sports league will actually reduce competitive balance. The result is driven by the so-called \textit{dulling effect}, which states that revenue sharing reduces the incentives for clubs to invest in playing talent because each club has to share some of the resulting marginal benefits of its talent investment with the other clubs in the league. Finally, Dietl et al. (2011) obtain, in a more general model assuming that club owners maximize an objective function given by a weighted sum of profits and winning percentage, that revenue sharing reduces or enhances investment incentives and deteriorates or improves competitive balance in the league, depending on whether 
the dulling effect or its opposite \textit{sharpening effect} occur.   
More recently, Moreno-Ternero et al. (2026) have studied the effect that revenue allocation schemes at international competition has on domestic competition. 

Textbook analyses of the economics underlying sportscasting, including
issues such as demand, supply, ownership, and market intervention have been
provided in the literature (e.g., Gratton and Solberg, 2007). The aim of
this survey is to provide an overview on the topic of sharing sportscast
revenues within theoretical economics, but we acknowledge that the topic has
also received attention in sports management and media economics. For
instance, some emphasis has been made on the comparative analysis of the
regulation of television sportscasting, suggesting that regulation should
actually seek to balance the commercial priorities of broadcasters and
sports organizations with the wider sociocultural benefits citizens gain
from free-to-air sportscasting (e.g., Smith et al., 2015). 
Another aspect receiving increasing attention in that literature is the
interaction effects of broadcasting different competitions worldwide. Thanks
to technology innovations, most of the world's population can watch sport
games from almost everywhere nowadays. Consequently, clubs are investing
heavily in promoting themselves to international markets, and networks
compete to broadcasting international sport competitions. Solberg and
Gaustad (2022) study the (mixed) success of major European and North
American competitions in this venture. Nalbantis et al., (2023) study the
transnational demand for European soccer telecasts and find sizeable
substitution effects, suggesting significant competition for the soccer
audience in the US market. 

The sportscast revenue allocation problem can also be approached from the perspective of fairness.
There is a sizable literature on fair allocation (e.g., Thomson, 2011) and
the concept has also been scrutinized in sports (e.g., Pawlenka, 2005; Csat%
\'{o}, 2021; Dietzenbacher and Kondratev, 2023). There exists consensus that
the revenue allocation in sports should reward performance 
thus guaranteeing incentive compatibility. But even this minimal goal is not
always achieved (e.g., Csat\'{o}, 2022, 2023).

Other aspects of the sportscasting industry, such as the role of
uncertainty of outcome and star quality in audiences (e.g., Buraimo and
Simmons, 2015), the strategies of small and new leagues (e.g., Fujak et al.,
2022; Tickell et al., 2024) or measuring audiences (e.g., Buraimo et al.,
2022; Oh and Kang, 2022; Fujak et al., 2024) have also been recently studied
within that literature. 

We should also acknowledge that, apart from sportscast revenues, other
sources of revenues are also important for sports organizations. This is,
for instance, the case of ticket sales and transfer fees, which have
received consirable attention in the literature 
(e.g., Feess and Muehlheusser, 2003; Hoey et al., 2021; Minchuk, 2024; van
Ours, 2024). That is why we shall not consider them here.

To conclude, we should mention that there exist two interesting recent surveys that could be seen as somewhat complmentary to this one. On the one hand, Devriesere et al., (2025) provide an overview of the tournament design literature from the perspective of operational research. Their emphasis is on the three main pillars of tournament design; namely, efficacy, fairness, and attractiveness.
On the other hand, Palacios-Huerta (2025) provides an impressive comprehensive set of economic techniques that can be tested using sports data, with the advantage that, in sports, features that often characterize either the lab or the field are found simultaneously.\footnote{In the words of Krumer and Szymanski (2025), ``It would not be an overstatement to say that this article is a manifesto for sports economics that could radically raise its status within economics.''} That survey emphasises what sports can do for economics. Our survey can actually be interpreted as an insight into the other direction, that is, what economic theory can do for the design of sports leagues.

The rest of the survey is organized as follows. In Section 2, we introduce
the problem focussing on the fan effect underlying the allocation process of sportscast revenues. 
In Section 3, we review the axiomatic approach to the problem. In Section 4,
we review the game-theoretical approach to the problem. In Section 5, we
review the decentralized approach via voting. In Section 6, we review
extensions of the model. 
In Section 7, we review some empirical applications. 
Finally, we conclude in Section 8.


\section{The fan effect}

Most of the contents of this section come from Berganti\~{n}os and
Moreno-Ternero (2020a). To motivate the problem we address in this survey,
we consider first the following example. 
A double round-robin tournament involves three clubs ($1$, $2$ and $3$) with
the following audiences: each of the (two) times $1$ plays $2$ the audience
is $1.2$ million; each of the (two) times $1$ plays $3$ the audience is $%
1.03 $ million; and each of the (two) times $2$ plays $3$ the audience is $%
0.23$ million. Overall (assuming each viewer pays $\$1$), the tournament
generates $\$4.92$ million. How should this amount be allocated among the
three clubs? To answer this question, note first that individuals watching a
game involving clubs $i$ and $j$ can be classified in four buckets:

\begin{enumerate}
\item Being a fan of this sport per se (in which case one would watch all
the games, independently of the clubs playing).

\item Being a fan of club $i$ (in which case one would watch all the games
involving club $i$).

\item Being a fan of club $j$ (in which case one would watch all the games
involving club $j$).

\item Considering that the game between clubs $i$ and $j$ is interesting.
\end{enumerate}

\bigskip

We can illustrate this situation by a triangle where the nodes are the
matches. Group 1 is at the centre of the triangle, Groups 2 and 3 are on the
sides, and Group 4 is on the nodes.

\bigskip

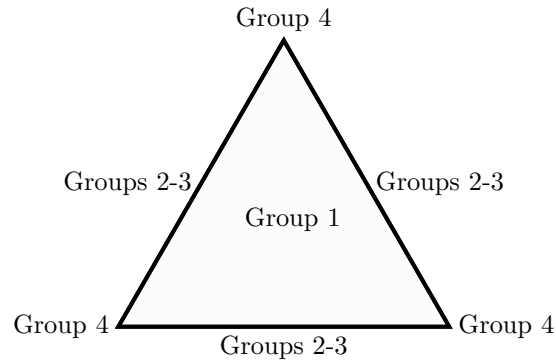
\begin{center}
	
	\begin{tikzpicture}[scale=1.1]
		
		\draw [line width=1.5pt, fill=gray!2] (0,0) -- (60:4) -- (4,0) -- cycle;
		
		\coordinate[label=left: Group 4]  (A) at (0,0);
		\coordinate[label=right:Group 4] (B) at (4,0);
		\coordinate[label=above: Group 4] (C) at (2,3.464);
		
		\coordinate[label=below: Groups 2-3](c) at ($ (A)!.5!(B) $);
		\coordinate[label=left: Groups 2-3] (b) at ($ (A)!.5!(C) $);
		\coordinate[label=right: Groups 2-3](a) at ($ (B)!.5!(C) $);
		
		\coordinate[label=left: Group 1]  (D) at (2.8,1.3);

		
		
	
	\draw [line width=1.5pt] (A) -- (B) -- (C) -- cycle;
\end{tikzpicture}

\end{center}

\bigskip

In practice, the information about whether one viewer belongs to one of the
four buckets is not available and only the total audience of the game is
known. Several plausible scenarios for the given audiences are presented
next.%
\medskip

\underline{Scenario $\left( a\right) $}. All viewers belong to group 1 and,
thus, game audiences are ignored. In this case, the revenue is equally
allocated among all participating clubs. That is, the allocation would be $%
(1.64,1.64,1.64)$. \medskip

\underline{Scenario $\left( b\right) $}. All viewers belong to group 4 and,
thus, no club has fans. In this case, it seems natural to divide viewers of
each game equally. That is, the allocation would be $(2.23,1.43,1.26)$. 
\medskip

\underline{Scenario $\left( c\right) $}. Club 1 has $1$ million fans, club 2
has $0.2$ million fans and club 3 has $0.03$ million fans. No viewers belong
to groups 1 or 4. In this case, it seems natural to concede each club the
amount generated by its fans. 
The allocation would then be $(4,0.8,0.12)$. 
\medskip

In general, consider a double round-robin tournament involving a finite set of
clubs $N=\{1,2,\dots ,n\}$. For each pair of clubs $i,j\in N$, let $a_{ij}$
denote the broadcasting audience (number of viewers) for the game played by $%
i$ and $j$ at $i$'s stadium. We use the notational convention that $a_{ii}=0$%
, for each $i\in N$. These numbers are collected in the \textit{audience
matrix} $A$. Let $\alpha _{i}=\sum_{j\in N}(a_{ij}+a_{ji})$ denote the
overall audience achieved by club $i$ throughout the tournament. 

For ease of exposition, the revenue generated from each viewer is normalized
to $1$ (to be interpreted as the \textquotedblleft pay per
view\textquotedblright\ fee). Thus, the overall revenues raised in the
tournament are given by 
$||A||=\sum_{i,j\in N}a_{ij}=\frac{1}{2}\sum_{i\in N}\alpha _{i}$. 
A \textbf{rule} is a mapping that associates with each problem the
allocation of the overall revenue among participating clubs.

The rather extreme scenario $\left( a\right) $ above gives rise to the
following rule, which splits evenly the overall revenue. 

\medskip \noindent \textbf{Uniform rule}, $U$: for each audience matrix $A$,
and each $i\in N$, 
\begin{equation*}
U_{i}(A) =\frac{\left\vert \left\vert A\right\vert \right\vert }{n}.
\end{equation*}

Scenarios $\left( b\right) $ and $\left( c\right) $ above can also be
thought of as extreme scenarios, but now only regarding the \textit{fan
effect}. They can be generalized as follows. In the first (\textit{minimalist%
}) scenario, it is assumed that there are no fans. Thus, it seems natural to
divide $a_{ij}$ equally between clubs $i$ and $j$, for each pair $i,j\in N$.
Formally,

\medskip \noindent \textbf{Equal-split rule}, $ES$: for each audience matrix 
$A$, and each $i\in N$, 
\begin{equation*}
ES_{i}(A) =\frac{\alpha _{i}}{2}=\frac{\sum_{j\in N}(a_{ij}+a_{ji})}{2}.
\end{equation*}


In the second (\textit{maximalist}) scenario, it is assumed there exist as
many fans as possible, compatible with the data. This is implemented by
minimizing the number of viewers in group 4.

Formally, for each pair of clubs $i,j\in N,$ with $i\neq j$, let 
\begin{equation*}
a_{ij}=b_{0}+b_{i}+b_{j}+\varepsilon _{ij},
\end{equation*}%
where $b_{0}$ denotes the number of generic sport fans, $b_{k}$ denotes the
number of fans of club $k=i,j$, and $\varepsilon _{ij}$ denotes the number
of joint fans for the pair $\{i,j\}$. 

Fix $k\in N$, and consider the following minimization problem: 
\begin{equation}
\min\limits_{b\in \mathbb{R}^{n}}\sum_{i,j\in N,i\neq j}\varepsilon _{ij}^{2},
\label{minim problem for N menos k}
\end{equation}%
where 
\begin{equation*}
\varepsilon _{ij}=\left\{ 
\begin{tabular}{cc}
$a_{ij}-b_{0}-b_{i}-b_{j}$ & if $k\notin \left\{ i,j\right\} $ \\ 
$a_{ij}-b_{0}-b_{i}$ & if $k=j$ \\ 
$a_{ij}-b_{0}-b_{j}$ & if $k=i$.%
\end{tabular}%
\right.
\end{equation*}

Let $\hat{b}_{0}$ and $\left\{ \hat{b}_{i}\right\} _{i\in N\backslash
\left\{ k\right\}}$ denote the solutions to $\left( \ref{minim problem for N
menos k}\right)$. 
Finally, for each pair $i,j\in N,$ with $i\neq j$, let 
\begin{equation*}
\hat{\varepsilon} _{ij}=\left\{ 
\begin{tabular}{cc}
$a_{ij}-\hat{b}_{0}-\hat{b}_{i}-\hat{b}_{j}$ & if $k\notin \left\{
i,j\right\} $ \\ 
$a_{ij}-\hat{b}_{0}-\hat{b}_{i}$ & if $k=j$ \\ 
$a_{ij}-\hat{b}_{0}-\hat{b}_{j}$ & if $k=i$.%
\end{tabular}%
\right.
\end{equation*}
\medskip

The following procedures are then considered to allocate $a_{ij}$:

$\left( P1\right) $ $\hat{b}_{0}$ is divided equally among all clubs.

$\left( P2\right) $ $\hat{b}_{l}$ is assigned to club $\ell$, for each $\ell\in
N\setminus\{k\}$.

$\left( P3\right) $ $\hat{\varepsilon}_{ij}$ is divided equally between
clubs $i$ and $j$, for each pair $i,j\in N,$ with $i\neq j$.

\bigskip

The above suggests the following rule. For each audience matrix $A$, and
each $i\in N$,

\begin{equation}  \label{regression rule formula}
R_{i}^{b,k}(A) =\left\{ 
\begin{tabular}{cc}
$\left( n-1\right) \widehat{b_{0}}+2\left( n-1\right) \widehat{b_{i}}%
+\sum_{j\in N\backslash \left\{ i\right\} }\frac{\widehat{\varepsilon _{ij}}+%
\widehat{\varepsilon _{ji}}}{2}$ & if $i\ne k $ \\ 
$\left( n-1\right) \widehat{b_{0}}+\sum_{j\in N\backslash \left\{ i\right\} }%
\frac{\widehat{\varepsilon _{ij}}+\widehat{\varepsilon _{ji}}}{2}$ & if $i=
k $%
\end{tabular}%
\right.
\end{equation}

\bigskip

One might argue that the above allocation would depend on $k$. This is not
the case. As stated in the next theorem, the allocation rule, so
constructed, coincides with the following rule, dubbed \textit{concede-and-divide}%
. 

\medskip \noindent \textbf{Concede-and-divide}, $CD$: for each audience
matrix $A$, and each $i\in N$, 
\begin{equation*}
CD_{i}(A) =\alpha _{i}-\frac{\sum\limits_{j,k\in N\backslash \left\{
i\right\} }\left( a_{jk}+a_{kj}\right) }{n-2}=\frac{\left( n-1\right) \alpha
_{i}-\left\vert \left\vert A\right\vert \right\vert }{n-2}.
\end{equation*}

\begin{theorem}
(Berganti\~{n}os and Moreno-Ternero, 2020a) \label{OLS well defined}For each
audience matrix $A$, and each pair $i,k\in N,$ let $R_{i}^{b,k}(A) $ be the
allocation obtained by applying formula $\left( \ref{regression rule formula}%
\right) $. Then, $R_{i}^{b,k}(A) =%
CD_{i}(A). $
\end{theorem}


Note that \textit{concede-and-divide} could also be rationalized in an
alternative way. For each club $i\in N$, the rule assigns its overall
audience throughout the tournament ($\alpha _{i}$) after conceding each of
the remaining clubs ($n-1$) the average audience of the games $i$ did not
play $\left(\frac{\sum\limits_{j,k\in N\backslash \left\{ i\right\} }\left(
a_{jk}+a_{kj}\right) }{(n-1)(n-2)}\right)$.

The \textit{equal-split rule} and \textit{concede-and-divide} therefore
provide polar estimates of the fan effect. Based on this, one might argue
that they should provide a range in which allocations 
should lie. In the example mentioned above, club 1 should receive something
between $2.23$ and $4$ million, club 2 something between $0.8$ and $1.43$
million and club 3 something between $0.12$ and $1.26$ million. For
instance, consider the following scenario:

\underline{Scenario $\left( d\right) $}. Club 1 has $0.8$ million fans, club
2 has $0.1$ million fans and club 3 has $0.03$ million fans. $0.09$ million
viewers belong to group 1. The remaining viewers belong to group 4. In this
case, it seems natural to concede each club the amount generated by its fans
and divide the rest equally. That is, the suggested allocation is $%
(3.7,0.8,0.42)$. 
\medskip

We shall present in the next section some rules implementing that goal.

\section{An axiomatic approach}

In this section, we review the axiomatic approach to derive specific rules
(such as those presented in the previous section and compromises among
them). That is, rather than proposing rules directly, the focus is on
formalizing axioms that reflect properties for those rules with a normative
appeal. Several combinations of some of those axioms will eventually lead
towards specific rules.

We start presenting two fundamental axioms with a long tradition in
axiomatic work. The first one is a structural axiom that says that revenues
should be additive on $A$. This notion can be traced back to Shapley (1953). 
Formally, \medskip

\noindent \textbf{Additivity}: 
For each pair of audience matrices $A$ and $A^{\prime }$, 
$R\left( A+A^{\prime }\right) =R(A)+R\left( A^{\prime }\right)$.\medskip

The second one models impartiality, a notion with a long tradition in the theory of justice (e.g.,
Moreno-Ternero and Roemer, 2006). 
It states that if two clubs have the same audiences each time they play a
third, then they should receive the same amount. Formally, \medskip

\noindent \textbf{Equal treatment of equals}: 
For each audience matrix $A$, and each pair $i,j\in N$ such that $%
a_{ik}=a_{jk}$, and $a_{ki}=a_{kj}$, for each $k\in N\setminus \{i,j\}$, $%
R_{i}(A)=R_{j}(A)$.\medskip

It turns out that these two axioms characterize the following family of
rules that arise from compromising between the \textit{uniform} rule and 
\textit{concede-and-divide} (both introduced in Ssection 2). 


\medskip \noindent \textbf{Generalized compromise rules} $\left\{
UC^{\lambda }\right\} _{\lambda \in \mathbb{R}}$: for each $\lambda \in 
\mathbb{R}$ each audience matrix $A$, and each $i\in N$, 
\begin{equation*}
UC_{i}^{\lambda }(A)=(1-\lambda )U_{i}(A)+\lambda CD_{i}(A).
\end{equation*}%
%
%
%
%
%
%
%
%
%

\begin{theorem}
\label{char ETE+AD} (Berganti\~{n}os and Moreno-Ternero, 2022a) A rule
satisfies \textit{additivity} and \textit{equal treatment of equals} if and
only if it is a generalized compromise rule.
\end{theorem}

A weaker version of \textit{equal treatment of equals} can also be
formalized, stating that if two clubs have the same audiences, not only when
facing each of the other clubs, but also when facing themselves at each
stadium, then they should receive the same amount. An axiom of \textit{%
pairwise reallocation proofness} (which says that a redistribution between
the audiences of the two games involving a pair of clubs does not affect the
revenues obtained by the clubs in the pair) fills the gap between both
axioms (e.g., Berganti\~{n}os and Moreno-Ternero, 2022a). Thus, parallel
characterization results to those listed here can be obtained upon replacing 
\textit{equal treatment of equals} by the combination of \textit{weak equal
treatment of equals} and \textit{pairwise reallocation proofness}. 
Similarly, if \textit{equal treatment of equals} is replaced by \textit{%
anonymity} (another impartiality axiom indicating that the name of the
agents does not matter) in Theorem \ref{char ETE+AD}, a more general family
of rules (that shall be discussed in Section 5), is characterized
(e.g., Berganti\~{n}os and Moreno-Ternero, 2023a). In this family, the amount
received by each club has three components: one depending on its (overall) home
audience, another depending on its (overall) away audience, and the third
depending on the overall audience in the whole tournament. Finally, adding
to the axioms of \textit{additivity} and \textit{equal treatment of equals}
(or \textit{anonymity}) natural axioms reflecting focal lower or upper
bounds in the allocations, natural subfamilies of the generalized compromise
rules, or extensions of them, are characterized (e.g., Berganti\~{n}os and
Moreno-Ternero, 2021, 2022a, 2024a; Zou and Mei, 2025). For instance, the family made of all the convex
combinations between the equal-split rule and concede-and-divide is characterized with the combination of additivity, equal treatment of equals, and \textit{maximum aspirations}, which says that each team should receive, at
most, the total audience of the games it played (e.g., Berganti\~{n}os and Moreno-Ternero, 2021).
Note that the \textit{equal-split} rule is a generalized compromise rule. As
a matter of fact, it is the only generalized compromise rule that satisfies
the basic axiom of \textit{null team}, which states that clubs with null
audience get a null reward. 
That is, a rule satisfies \textit{additivity}, \textit{equal treatment of
equals}, and \textit{null team} if and only if it is the \textit{equal-split
rule}. 
On the other hand, \textit{concede-and-divide} is the only compromise rule
that satisfies the counterpart axiom of \textit{essential team} (e.g.,
Berganti\~{n}os and Moreno-Ternero, 2020a), which states that if only the
games played by one team may have positive audience, 
then such an essential team should receive all its audience.

The combination of \textit{null team} with \textit{anonymity} and additivity \textit{anonymity} characterizes 
natural generalizations of the \textit{equal-split rule}.
To wit, for each $\lambda \in \mathbb{R}$ and each game $\left( i,j\right) $%
, $S^{\lambda }$ divides the audience $a_{ij}$ among the clubs $i$ and $j$
proportionally to $\left( 1-\lambda ,\lambda \right) $. Formally, for each
audience matrix $A$ and each $i\in N,$ 
\begin{equation*}
S_{i}^{\lambda }\left( A\right) =\sum_{j\in N\backslash \left\{ i\right\}
}\left( 1-\lambda \right) a_{ij}+\sum_{j\in N\backslash \left\{ i\right\}
}\lambda a_{ji}.
\end{equation*}

The \textit{equal-split rule} corresponds to the case where $\lambda =0.5.$
When $\lambda =0$ all the audience is assigned to the home club and when $%
\lambda =1$ all the audience is assigned to the away club. We say that $R$
is a \textbf{split rule} if $R\in \left\{ S^{\lambda }:\lambda \in \left[ 0,1%
\right] \right\} .$ We say that $R$ is a \textbf{generalized split rule} if $%
R\in \left\{ S^{\lambda }:\lambda \in \mathbb{R}\right\} .$

\begin{theorem}
(Berganti\~{n}os and Moreno-Ternero, 2024a) \label{char gen MS} 
A rule satisfies \textit{additivity}, \textit{anonymity}, and \textit{null
team} if and only if it is a generalized split rule. 
\end{theorem}

The \textit{split rules} are characterized when adding to the previous
statement one of the focal bounds mentioned above, or a \textit{monotonicity}
axiom (e.g., Berganti\~{n}os and Moreno-Ternero, 2022b, 2022c, 2024a).%
\medskip



We conclude this section mentioning two caveats for all the results presented above.

On the one hand, they all make use of \textit{additivity}. Alternative axioms formalizing the notion
of \textit{marginalism} have also been considered, giving rise to several
characterizations of some of the rules introduced above (e.g., Berganti\~{n}os and Moreno-Ternero, 2020b).

On the other hand, they all assume that the population $%
\left( N\right) $ is fixed. It seems natural to obtain characterizations for
a variable-population setting. Two classical variable-population axioms are 
\textit{consistency} and \textit{population monotonicity} (e.g. Thomson,
2011). The former states that for each problem and for the \textquotedblleft
reduced problem\textquotedblright\ (obtained by imagining the departure of a
group of agents with their allocation, and reassessing the remaining amount
to the remaining agents) the allocation proposed by the rule (for the
subgroup) is the same. The latter states that if a new club joins a league,
then no club from the initial league is worse off. It would be interesting
to explore the implications of those axioms. These are open questions for
future research.

\section{A game-theoretical approach}

A natural course of action in economics is to solve problems indirectly, via
associating them to cooperative games for which well-established solutions
are available. Formally, a \textbf{cooperative game with transferable utility%
}, briefly a \textbf{TU game}, is a pair $\left( N,v\right) $, where $N$
denotes a set of agents and $v:2^{N}\rightarrow \mathbb{R}$ satisfies $%
v\left( \varnothing \right) =0.$ As the population $N$ will remain fixed, we
avoid its use in the notation.

The \textbf{core} is defined as the set of feasible payoff vectors, upon
which no coalition can improve. Formally, 
\begin{equation*}
Core\left( v\right) =\left\{ x\in \mathbb{R}^{N}\text{ such that }\sum_{i\in
N}x_{i}=v\left( N\right) \text{ and }\sum_{i\in S}x_{i}\geq v\left( S\right) 
\text{, for each }S\subset N\right\} .
\end{equation*}

The \textbf{Shapley value} (Shapley, 1953) is defined for each player as the
average of his contributions across orders of agents. Formally, for each $%
i\in N$, 
\begin{equation*}
Sh_{i}\left( v\right) =\frac{1}{n!}\sum_{\pi \in \Pi _{N}}\left[ v\left(
Pre\left( i,\pi \right) \cup \left\{ i\right\} \right) -v\left( Pre\left(
i,\pi \right) \right) \right] ,
\end{equation*}%
where $\Pi _{N}$ denotes the set of all orders on $N$, and $Pre\left( i,\pi
\right) =\left\{ j\in N\mid \pi \left( j\right) <\pi \left( i\right)
\right\} $.

The \textbf{egalitarian value} (e.g., van den Brink, 2007) yields each agent
an equal portion of the value of the grand coalition. Formally, for each $%
i\in N$, 
\begin{equation*}
ED_{i}\left( v\right) =\frac{1}{n}v(N).
\end{equation*}

The \textbf{egalitarian Shapley values} (e.g., Casajus and Huettner, 2013) 
are obtained with the convex combinations of the previous values. Formally,
for each $i\in N$ and each $\lambda \in \left[ 0,1\right]$, 
\begin{equation*}
S_{i}^{\lambda }\left( v\right) =\lambda Sh_{i}\left( v\right) +(1-\lambda
)ED_{i}\left( v\right) .
\end{equation*}


Berganti\~{n}os and Moreno-Ternero (2020a) associate with each sportscast
problem 
a TU game. 
To do so, they take an optimistic stance on what revenue a coalition can
generate on its own. To wit, the highest possible revenue that a game
between clubs $i$ and $j$ in the former's stadium may generate is $a_{ij}$.
Thus, by breaking away from the league, the most optimistic scenario for any
coalition of clubs is to generate the same revenue they generated before
leaving the league. Formally, for each $S\subset N,$ $v_{A}\left( S\right) $
is defined as the total audience of the games played by the clubs in $S$.
Namely, 
\begin{equation*}
v_{A}\left( S\right) =\sum_{i,j\in S,i\neq j}a_{ij}=\sum_{i,j\in
S,i<j}\left( a_{ij}+a_{ji}\right) .
\end{equation*}


Berganti\~{n}os and Moreno-Ternero (2020a) show the correspondence between
the \textit{equal-split rule} and the Shapley value of the corresponding TU
game. Now, van den Nouweland et al., (1996) studied the so-called Terrestial
Flight Telephone System via a formally equivalent game to the one mentioned
here. 
They proved that, 
in this game, the Shapley value coincides with the Nucleolus and the $\tau $%
-value. Thus, all these values collapse into the \textit{equal-split rule}
for this setting, which reinforces this rule from a game-theoretical
approach.

Berganti\~{n}os and Moreno-Ternero (2022a) argue that there is
also a correspondence between the \textit{uniform rule} and the equal
division value of the corresponding TU game. Consequently, there is also a
correspondence between the rules that compromise between the \textit{uniform
rule} and the \textit{equal-split rule} (that is, the rules that belong to
the \textit{generalized compromise rules} for $\lambda \in \left[ 0,\frac{n-2%
}{2\left( n-1\right) }\right] $) and the egalitarian Shapley values of the
corresponding TU game. Gon\c{c}alves-Dosantos et al., (2022) have recently
introduced a new value in cooperative games. This value is characterized
with an axiom called \textit{necessary players}, which is reminiscent of the 
\textit{essential team} axiom mentioned above. It remains an open question
to explore whether there is a connection between \textit{concede-and-divide}
and such a value (or related ones). Similarly, the connections between other
families of rules in sportscast problems and families of values in
cooperative games have not been explored yet.

Berganti\~{n}os and Moreno-Ternero (2020a) also show that, in order to
satisfy the core constraints (of the game $v_{A}$), one should divide the
revenue generated by the audience of a game between the two clubs playing
the game. As the Shapley value coincides with the \textit{equal-split rule}, in which the revenue generated by the audience of a game is divided equally between the two clubs playing the game, this implies that the Shapley value is always within the core. 
Berganti\~{n}os and Moreno-Ternero (2024b) go further to show that
the allocations within the core can be obtained as the allocations induced
by the set of rules satisfying additivity, null team and a monotonicity
axiom. 

\section{A decentralized approach}

The previous sections gathered normative and positive foundations for many
rules to share sportscast revenues. Nevertheless, no strong consensus exists
over specific rules (albeit two of them; namely \textit{equal-split} and 
\textit{concede-and-divide} seem to be salient). This motivates to explore a
different (decentralized) approach, following a long tradition of allocating
resources by voting (e.g., Birnberg et al., 1970; Barzel and Sass, 1990), in
which the choice of a rule could be made by means of simple majority voting,
letting each club vote for a rule.%

Given an audience matrix $A$, $R\left( A\right) $ is a \emph{majority winner}
(within the set of rules $\mathcal{R}$) for $A$ if there is no other rule $%
R^{\prime }\in \mathcal{R}$ such that $R_{i}^{\prime }(A)>R_{i}(A)$ for a
majority of clubs. The family of rules $\mathcal{R}$ has a \emph{majority
voting equilibrium} if there is at least one majority winner (within $%
\mathcal{R}$) for each audience matrix. 

Berganti\~{n}os and Moreno-Ternero (2023a) show that there is no majority voting equilibrium for the family of rules satisfying additivity and anonymity and, thus, a Condorcet cycle (paradox) exists. 
The underlying rationale
is that given one of those rules, one can construct another which, at a
certain problem, increases the amount obtained by a majority of the clubs
involved, while reducing the amount obtained by all of the others. 
Nevertheless, the existence of a majority voting equilibrium is guaranteed
for other sufficiently large subfamilies of rules. In particular, for the
family of generalized compromise rules (i.e., as mentioned in Section 3,
those rules satisfying additivity and equal treatment of equals). This is a
consequence of the fact that those rules satisfy the so-called \textit{%
single-crossing} property (e.g., Berganti\~{n}os and Moreno-Ternero 2023a).
That is, for each pair of rules within the family, and each problem, there
exists a club separating those clubs benefitting from the choice of one rule
and those benefitting from the choice of the other. It is well known that
the single-crossing property of preferences is a sufficient condition for
the existence of a majority voting equilibrium (Gans and Smart 1996). Thus,
the next result follows.

\begin{theorem}
(Berganti\~{n}os and Moreno-Ternero, 2023a). \label{eq} There is a majority
voting equilibrium for each bounded family of generalized compromise rules $%
\left\{ UC^{\lambda }\right\} _{\lambda \in \lbrack \underline{\lambda },%
\overline{\lambda }\rbrack}$.
\end{theorem}

Theorem \ref{eq} states that if we let clubs vote for a rule within any
bounded family of generalized compromise rules, then there will be a
majority winner for each problem. The identity of this winner will be
problem specific and it will depend on the characteristics of the problem at
stake. In most cases, either the \textit{equal-split rule} or \textit{%
concede-and-divide} arise, but more rules could also arise (e.g., Berganti\~{n}os and Moreno-Ternero, 2023a).

Another consequence of the single-crossing property is that it guarantees
progressivity comparisons of schedules (Jakobsson 1976; Hemming and Keen
1983). Thus, one can also obtain an interesting result, referring to the
distributive power of the rules within the family of generalized compromise
rules. 
Formally, given $x,y\in \mathbb{R}^{n}$ satisfying $x_{1}\leq x_{2}\leq
...\leq x_{n}$, $y_{1}\leq y_{2}\leq ...\leq y_{n}$, and $%
\sum_{i=1}^{n}x_{i}=\sum_{i=1}^{n}y_{i}$, we say that $x$ \textit{is greater
than }$y$ \textit{in the Lorenz ordering} if $\sum_{i=1}^{k}x_{i}\geq
\sum_{i=1}^{k}y_{i}$, for each $k=1,...,n-1$, with at least one strict
inequality. When $x$ is greater than $y$ in the Lorenz ordering, one can
state (see, for instance, Dasgupta et al., 1973) that $x$ is unambiguously
\textquotedblleft more egalitarian\textquotedblright\ than $y$. In our
setting, we say that a rule $R$ \textit{Lorenz dominates} 
another rule $R^{\prime }$ if for each audience matrix $A$, $R(A)$ is
greater than $R^{\prime }(A)$ in the Lorenz ordering. As the Lorenz
criterion is a partial ordering, one might not expect to be able to perform
many comparisons of vectors. It turns out, however, that 
the generalized compromise rules are fully ranked according to this
criterion. 

\begin{proposition}
(Berganti\~{n}os and Moreno-Ternero 2023a). \label{egalitarian UC} The
following statements hold:

\begin{itemize}
\item If $0\le\lambda _{1}\leq \lambda _{2}$ then $UC^{\lambda _{1}}$ Lorenz
dominates $UC^{\lambda _{2}}$.

\item If $\lambda _{1}\leq \lambda _{2} \le 0$ then $UC^{\lambda _{2}}$
Lorenz dominates $UC^{\lambda _{1}}$.
\end{itemize}
\end{proposition}

Proposition \ref{egalitarian UC} implies that the parameter defining the
family can actually be interpreted as an index of the distributive power of
the rules within the family. The \textit{uniform rule} is the center of the
family, obtained when $\lambda= 0$. It also happens to be the \textit{maximal%
} element of the Lorenz ordering, as it generates fully egalitarian
allocations. It is then obvious that all other rules within the family are
Lorenz dominated by it. The remarkable feature, that Proposition \ref%
{egalitarian UC} states, is that departing from the \textit{uniform rule} in
both directions (either with positive parameters or with negative
parameters) one obtains rules that yield progressively less egalitarian
allocations. That is, the more one departs from the center, the less
egalitarian rules become. And those comparisons can be established for each
pair of rules within each of the two sides of the family. When the pair of
rules is made of rules in different sides of the family (i.e., one
corresponding to a negative parameter and the other corresponding to a
positive parameter) then one cannot establish Lorenz comparisons for such a
pair of rules.

\section{Cancelled seasons and the operational approach}

The above model has to be extended to accommodate the case where a league
has been cancelled. This was particularly relevant in the aftermath of the
COVID-19 pandemic, which forced the partial or total cancellation of many
sports competitions worldwide. The reader is referred to Alabi and Urquhart
(2023) for a detailed account of the impact of the COVID-19 pandemic on top
English football clubs. The question that arises is how the allocation of
the sportscast revenues should be modified in such a scenario. 
Although we frame the analysis for cancelled leagues, the model we present in this section would also cover the case of competitions with incomplete round-robin formats, which are increasingly frequent (for instance, in UEFA club competitions). The reader is referred to Devriesere and Goossens (2025) and Li et al. (2025) for recent analyses of this case. 

In the benchmark sportscast problem (with no cancellations) considered
above, a simplifying assumption is made. Instead of dividing the revenue
from the sale of broadcasting rights, the total audience of the season $%
\left( \left\vert \left\vert A\right\vert \right\vert =\sum_{i,j\in
N}a_{ij}\right)$ is divided. Then, the theoretical model only considers
audience matrix $A$ as the input of the problem. In the model of the
sportscast problem with cancellations, there are two inputs: $A$, the
audience matrix (in which some entries could be empty) and $E$, the amount
to divide among the clubs. Consider the following examples from Berganti\~{n}%
os and Moreno-Ternero (2023c).

\begin{example}
\label{example motivating estimation}Let $N=\left\{ 1,2,3,4\right\} ,$ $%
E=100,$ and 
\begin{equation*}
A=\left( 
\begin{array}{cccc}
\varnothing & \varnothing & 10 & 10 \\ 
10 & \varnothing & 1 & 1 \\ 
10 & 1 & \varnothing & \varnothing \\ 
10 & 1 & 1 & \varnothing%
\end{array}%
\right) \text{. }
\end{equation*}
In this case, a (double round-robin) league with four clubs is considered,
where the last round of the tournament (games 1-2 and 3-4) was cancelled.
Furthermore, one of the clubs in the league (club 1) is stronger (in terms
of audiences) than the other three (normal) clubs that are symmetric. The
goal is to divide the generated revenue (100) among the clubs using the
information from matrix $A.$

In the case without cancellations, the equal-split rule divides the revenue
proportionally to the audiences of the clubs. It seems reasonable to apply
the same idea in this example. If so, one obtains the following: 
\begin{equation*}
\begin{tabular}{ccccc}
& Club 1 & Club 2 & Club 3 & Club 4 \\ 
Audiences & 50 & 14 & 23 & 23 \\ 
Allocation & 45.5 & 12.7 & 20.9 & 20.9%
\end{tabular}%
\end{equation*}

The above is equivalent to assign a 0 audience to the cancelled games and
compute the rule in the induced problem without cancellations.

One might wonder whether assigning a $0$ audience to all cancelled games is
the best option. Another reasonable option, for instance, would be to treat
all cancelled games equally. But this could be \textquotedblleft
unfair\textquotedblright\ for some clubs. To wit, in this example, when club
1 plays, the audience is $10$ but when club 1 does not play the audience is $%
1$. If one would assign an audience to the cancelled games, then it seems
more reasonable to assign an audience of $10$ to the cancelled game between
clubs $1$ and $2$ and an audience of $1$ to the cancelled game between clubs 
$3$ and $4$. Suppose now that one divides the revenue proportionally to the
new assigned audiences. Then, one obtains the following: 
\begin{equation*}
\begin{tabular}{ccccc}
& Club 1 & Club 2 & Club 3 & Club 4 \\ 
Assigned audiences & 60 & 24 & 24 & 24 \\ 
Allocation & 45.5 & 18.2 & 18.2 & 18.2%
\end{tabular}%
\end{equation*}

Given the configuration of this league, one could argue that club 2 is
unlucky because it loses the audience from a game against the strongest
club, whereas clubs 3 and 4 simply lost the audience of a game against a
\textquotedblleft normal" club. Thus, it seems fair to treat clubs 2, 3 and
4 equally, instead of rendering them contingent on the lottery on how
tournament rounds are designed.%
\end{example}



\begin{example}
\label{example motivating rules} Let $N=\left\{ 1,2,3,4\right\} ,$ $E=100,$
and 
\begin{equation*}
A=\left( 
\begin{array}{cccc}
\varnothing & 8 & 10 & 9 \\ 
10 & \varnothing & \varnothing & \varnothing \\ 
12 & \varnothing & \varnothing & \varnothing \\ 
11 & \varnothing & \varnothing & \varnothing%
\end{array}%
\right) \text{. }
\end{equation*}
In this case, a league with four clubs is considered, where only the games
involving club 1 have been played.

Assume that no viewer is a fan (or, in the parlance of Section 2, they all
belong to bucket 4), as in the case of the equal-split rule. Then, one
should divide the revenue proportionally to the audience of the clubs. Thus, 
\begin{equation*}
\begin{tabular}{ccccc}
& Club 1 & Club 2 & Club 3 & Club 4 \\ 
Audiences & 60 & 18 & 22 & 20 \\ 
Allocation & 50 & 15 & 18.3 & 16.7%
\end{tabular}%
\end{equation*}

Assume now that there are as many fans as possible (or, in the parlance of
Section 2, bucket 4 is empty), as in the case of concede-and-dvide. One
could then assume that club 1 has fans whereas the other clubs do not.
Notice that this explanation is compatible with the data (and the assumption
that no viewer belongs to bucket 4). Thus, club 1 receives 100 and the rest
of the clubs receive 0, which is in stark contrast with the previous
allocation.
\end{example}


The above examples illustrate two possible ways to solve sportscast problems
for cancelled seasons. First, to allocate the revenue using only the real
audiences from the non-cancelled games (or, equivalently, to assign $0$
audience to the cancelled games). Second, \textquotedblleft to
estimate\textquotedblright\ the audiences of the cancelled games and to
allocate the revenue using both the real audiences from the non-cancelled
games and the \textquotedblleft estimated\textquotedblright\ audiences of
the cancelled games. Several plausible options can be considered for this
estimation. One is to take as a proxy of the audience of a game between two
clubs, the audience of the game between the same clubs in the first leg of
the tournament.

An \textbf{extension operator} is defined via a mapping assigning to each
problem with possible empty entries in the audience matrix a benchmark
problem without any empty entries, with the proviso that non-empty entries
in the original matrix of audiences remain unchanged. The concept of
operators on the space of allocation rules is introduced by Thomson
and Yeh (2008) and explored further by Hougaard et al., (2012) and
Moreno-Ternero and Vidal-Puga (2021), among others. Two extension operators arise
naturally. First, the one associating to a cancelled game a zero audience.
Second, the one associating to a cancelled game the audience of the game in
the first leg of the tournament, or zero if such a game was also cancelled.
Formally,

\medskip \noindent \textbf{Zero}, $z$: For each pair $i,j\in N$, 
\begin{equation*}
a_{ij}^{z}=\left\{ 
\begin{tabular}{ll}
$0$ & if $a_{ij}=\varnothing $ \\ 
$a_{ij}$ & if $a_{ij}\neq \varnothing .$%
\end{tabular}%
\right.
\end{equation*}

\medskip \noindent \textbf{Leg}, $\ell $: For each pair $i,j\in N$, 
\begin{equation*}
a_{ij}^{\ell }=\left\{ 
\begin{tabular}{ll}
$a_{ji}$ & if $a_{ij}=\varnothing $ and $a_{ji}\neq \varnothing $ \\ 
0 & if $a_{ij}=\varnothing $ and $a_{ji}=\varnothing $ \\ 
$a_{ij}$ & if $a_{ij}\neq \varnothing .$%
\end{tabular}%
\right.
\end{equation*}%
\medskip

For each operator $o$, and each benchmark rule $R$, one can define an
extended rule $R^{o}$ in the obvious way. Some instances are the \textbf{%
zero-extended equal-split rule} ($ES^{z}$), the \textbf{zero-extended
concede-and-divide} ($CD^{z}$), the \textbf{leg-extended equal-split rule} ($%
ES^{\ell }$), and the \textbf{leg-extended concede-and-divide} ($CD^{\ell }$%
). Berganti\~{n}os and Moreno-Ternero (2023c) provide several
characterization results for these rules. 

\section{Empirical illustrations}

Although the contents presented in the previous sections are of a
theoretical nature, they can obviously be applied to real-life cases. In this section, we review the empirical
applications Berganti\~{n}os and Moreno-Ternero (2020a, 2021, 2023b) present, resorting to data from La Liga (the Spanish Football League), a double round-robin tournament involving 20 clubs.\footnote{\url{http://www.laliga.es/en}} 

\begin{table}[]
\begin{tabular}{lrrrrrrr}
\textbf{Club}             & \textbf{Audiences} & \textbf{Allocation} & \textbf{ES} & \textbf{CD}  & \textbf{\% Alloc.} & \textbf{\% ES} & \textbf{\% CD} \\
Real Madrid      & 43.610    & 140.10       & 152.32   & 252.29   & 11.24     & 12.22 & 20.23 \\
Barcelona        & 40.040    & 146.20       & 139.85   & 225.97   & 11.73     & 11.22 & 18.12 \\
Atl\'{e}tico de Madrid  & 23.210    & 99.20        & 81.07    & 101.87   & 7.96      & 6.50  & 8.17  \\
Betis            & 20.540    & 49.20        & 71.74    & 82.18    & 3.95      & 5.75  & 6.59  \\
Sevilla          & 19.260    & 65.90        & 67.27    & 72.75    & 5.29      & 5.40  & 5.83  \\
Real Sociedad    & 18.140    & 53.50        & 63.36    & 64.49    & 4.29      & 5.08  & 5.17  \\
Las Palmas       & 16.860    & 44.00        & 58.89    & 55.05    & 3.53      & 4.72  & 4.42  \\
M\'{a}laga           & 16.820    & 55.60        & 58.75    & 54.75    & 4.46      & 4.71  & 4.39  \\
Athletic Bilbao  & 15.980    & 71.00        & 55.81    & 48.56    & 5.69      & 4.48  & 3.89  \\
Villarreal        & 15.470    & 60.90        & 54.03    & 44.80    & 4.88      & 4.33  & 3.59  \\
Valencia         & 15.020    & 67.40        & 52.46    & 41.48    & 5.41      & 4.21  & 3.33  \\
Espa\~ nol          & 14.880    & 48.90        & 51.97    & 40.45    & 3.92      & 4.17  & 3.24  \\
Celta            & 13.540    & 51.40        & 47.29    & 30.57    & 4.12      & 3.79  & 2.45  \\
Deportivo La Coru\~ na & 13.420    & 44.00        & 46.87    & 29.68    & 3.53      & 3.76  & 2.38  \\
Granada          & 13.080    & 44.60        & 45.68    & 27.18    & 3.58      & 3.66  & 2.18  \\
Eibar            & 12.600    & 41.70        & 44.01    & 23.64    & 3.34      & 3.53  & 1.90  \\
Osasuna          & 11.440    & 43.00        & 39.96    & 15.08    & 3.45      & 3.20  & 1.21  \\
Legan\'{e}s          & 11.180    & 39.30        & 39.05    & 13.17    & 3.15      & 3.13  & 1.06  \\
Sporting         & 11.180    & 41.70        & 39.05    & 13.17    & 3.34      & 3.13  & 1.06  \\
Alav\'{e}s           & 10.710    & 39.30        & 37.41    & 9.70     & 3.15      & 3.00  & 0.78\\
                    &               &       &           &       &       
\end{tabular}

\textbf{Table 1}. La Liga 2016/2017. Actual allocation versus proposed allocations based on the equal-split rule and concede-and-divide.
\end{table}

First, Berganti\~{n}os and Moreno-Ternero (2020a) compare the allocation
implemented in the 2016-2017 season  with the ones obtained by applying the
equal-split rule or concede-and-divide. The results are presented in Table 1, where (as in the remaining tables) the audiences and the allocations are measured in millions
(viewers and euros, respectively), and clubs are ordered according to audiences. 

Several insights can be derived from Table 1. Contrary to what some
argue, La Liga's allocation seems to be biased against the
two powerhouses (Barcelona and Real Madrid). Although the \textit{%
equal-split rule} would jointly recommend for them a somewhat similar allocation (close to one fourth of the pie), \textit{concede-and-divide} would
recommend more (almost two fifths of the pie). Eight clubs are favored by
the actual allocation, in the sense that the amount they get is above the
amounts suggested by the two rules. Seven clubs obtain amounts between those
suggested by the two rules. Five clubs obtain amounts below those suggested
by the two rules.  

Finally, as the distribution of audiences is skewed to the left (more than half of the clubs have audiences strictly below the average) in this case, we know that for each bounded family of generalized compromise rules $%
\left\{ UC^{\lambda }\right\} _{\lambda \in \lbrack \underline{\lambda },%
\overline{\lambda }\rbrack}$ the majority voting equilibria is $UC^{\underline{%
\lambda }}.$ Thus, if, for instance, $[\underline{\lambda },\overline{\lambda }]=\left[ 0,1%
\right] $, then the uniform rule would the majority voting equilibrium.

\bigskip

Berganti\~{n}os and Moreno-Ternero (2021) switch towards the 2017-2018 season 
and apply therein the subset of compromise rules given by the convex combination of the \textit{equal-split}
rule and \textit{concede-and-divide}. Namely, $\lambda ES+\left( 1-\lambda
\right) CD$ where $\lambda \in \left[ 0,1\right] $. Note that, for this case, as a consequence of what we mentioned in the previous paragraph, 
the equal-split rule would the majority voting equilibrium.

In general, individuals
watching a game can be classified as fans of one of the clubs involved in
the game, or as \textit{neutral} viewers. In practice, the above information
is not available and one only knows the total audience of the game. Thus, $%
\lambda $ can be considered as an estimation of the percentage of \textit{%
neutral} viewers. Similarly, $1-\lambda $ can be considered as an estimation
of the percentage of viewers who watch a game because they are fans of one
of the clubs playing the game. It is argued that the amount received by each
club should be between the allocations proposed by the \textit{equal-split}
rule and \textit{concede-and-divide.}

Table 2 
shows the implemented allocation for the 2017-2018 season, and the
ones proposed by the two rules. In the last column, it is checked whether the
amount obtained by each club in the allocation used in practice corresponds
to some generalized compromise rule. For instance, Barcelona receives the
amount that the rule $0.98ES+0.02CD\ $(namely, $\lambda =0.98)$ would yield
for this setting. In contrast, Real Madrid receives less than the amount
proposed by any rule within the family ($148<\min \left\{
158.43,260.81\right\} )$ whereas Atl\'{e}tico de Madrid receives more ($%
110.60>\max \left\{ 85.77,107.43\right\} )$.
Nine clubs are favored by the actual allocation; namely, the amount each
gets is above the amounts suggested by both rules. Real Madrid and Betis
obtain amounts below those two rules. The remaining nine clubs obtain
amounts between both rules. However, the parameter $\lambda $ would be
different for each club. For instance, for Celta, it would be the rule
corresponding to $\lambda =0.02$ (something quite similar to the \textit{%
concede-and-divide} outcome), whereas, for Barcelona, it would be the rule
corresponding to $\lambda =0.98$ (something quite similar to the \textit{%
equal-split} outcome).

\begin{table}[]
\begin{tabular}{lrrrl}
\textbf{Club }            & \textbf{Allocation} & \textbf{ES}     & \textbf{CD}     & \textbf{$\lambda$} \\
Real Madrid      & 148.00       & 158.43 & 260.81 & Below  \\
Barcelona        & 154.00       & 151.70 & 246.61 & 0.98   \\
Betis            & 52.90        & 94.18  & 125.18 & Below  \\
Atl\'{e}tico Madrid  & 110.60       & 85.77  & 107.43 & Above  \\
Valencia         & 65.70        & 65.59  & 64.82  & Above  \\
Sevilla          & 74.00        & 62.23  & 57.72  & Above  \\
Celta            & 52.90        & 59.87  & 52.75  & 0.02   \\
M\'{a}laga           & 53.50        & 59.20  & 51.33  & 0.28   \\
Athletic Bilbao  & 73.20        & 57.85  & 48.49  & Above  \\
Espa\~ nol          & 52.40        & 56.17  & 44.94  & 0.66   \\
Las Palmas       & 46.80        & 53.48  & 39.26  & 0.53   \\
Levante          & 45.10        & 50.79  & 33.58  & 0.67   \\
Girona           & 61.50        & 50.12  & 32.16  & Above  \\
Real Sociedad    & 43.30        & 50.12  & 32.16  & 0.62   \\
Deportivo La Coru\~ na & 46.00        & 48.10  & 27.90  & 0.90   \\
Villarreal        & 65.50        & 46.42  & 24.35  & Above  \\
Alav\'{e}s           & 46.10        & 46.08  & 23.64  & Above  \\
Getafe           & 44.50        & 45.41  & 22.22  & 0.96   \\
Eibar            & 46.30        & 44.06  & 19.38  & Above  \\
Legan\'{e}s          & 43.30        & 40.03  & 10.86  & Above  \\
                 &              &        &        &        \\        
\end{tabular}

\textbf{Table 2}. La Liga 2017/2018. Actual allocation versus proposed
allocations based on the equal-split rule and concede-and-divide, as well as rationalization via compromise rules.
\end{table}


Berganti\~{n}os and Moreno-Ternero (2023b) also study the allocation of
revenues for La Liga, which is strongly regulated by the Spanish government
since 2015. More precisely, the Royal Decree published then decomposes $E$ (the revenue to
be allocated) in four parts, each reflecting a different dimension. The
amount received by each club is the sum of the amounts received in each
dimension. The four dimensions and the alternatives considered are described
next:

\begin{enumerate}
\item \textbf{Lower bounds}. Half of the total endowment is devoted to this
(first) dimension. It is divided equally among all clubs, hence guaranteeing
a specific \textit{lower bound} to each: $\frac{E}{40}$.

Berganti\~{n}os and Moreno-Ternero (2023b) propose two other lower bounds, inspired by the literature on claims problems
(e.g., Thomson, 2019).

\item \textbf{Sport performance}. One quarter of the total endowment is
devoted to this (second) dimension. It is divided among clubs taking into
account the sport performance during last five seasons.

Berganti\~{n}os and Moreno-Ternero (2023b) propose two alternatives: the Premier League's scheme, and awarding each club a score equal to the points obtained each season.

\item \textbf{Economic performance}. One twelfth of the total endowment is
devoted to this (third) dimension. It is divided among clubs proportionally
to ticket sales in the last five seasons.

Berganti\~{n}os and Moreno-Ternero (2023b) propose three alternatives, also inspired from the literature on on claims problems (as in the first dimension): 
the so-called constrained equal awards, constrained equal losses, and Talmud rules.

\item \textbf{Broadcasting performance}. One sixth of the total endowment is
devoted to this (last) dimension. The Royal Decree does not specify the way
in which this amount should be divided among clubs. La Liga decides the
amount received by each club, but it does not specify how such amounts are
computed.

Berganti\~{n}os and Moreno-Ternero (2023b) make proposals based on the equal-split rule and concede-and-divide,
respectively.
\end{enumerate}

\begin{table}[]
\begin{tabular}{lrrrrr}
\textbf{Club }            & \textbf{Allocation}  & \textbf{Low SD} & \textbf{High SD} & \textbf{Average} & \textbf{Difference} \\
Real Madrid         & 148.0          & 98.83  & 212.09  & 154.32  & -6.32      \\
Barcelona           & 154.0          & 99.73  & 210.75  & 154.24  & -0.24      \\
Betis               & 52.9         & 69.81  & 78.86   & 71.34   & -18.44     \\
Atl\'{e}tico de Madrid     & 110.6        & 85.37  & 122.93  & 102.02  & 8.58       \\
Valencia            & 65.7         & 74.56  & 69.24   & 69.58   & -3.88      \\
Sevilla             & 74.0          & 75.07  & 67.46   & 71.86   & 2.14       \\
Celta               & 52.9         & 62.23  & 49.38   & 56.48   & -3.58      \\
M\'{a}laga              & 53.5         & 61.12  & 44.76   & 54.52   & -1.02      \\
Athletic Bilbao     & 73.2         & 71.68  & 69.26   & 71.24   & 1.96       \\
Espa\~ nol             & 52.4         & 67.23  & 44.27   & 56.32   & -3.92      \\
Las Palmas          & 46.8         & 55.38  & 36.35   & 46.07   & 0.73       \\
Levante             & 45.1         & 55.01  & 35.21   & 45.14   & -0.04      \\
Real Sociedad       & 61.5         & 67.39  & 43.12   & 58.04   & 3.46       \\
Girona              & 43.3         & 48.41  & 33.70    & 41.48   & 1.82       \\
Deportivo La Coru\~ na & 46.0           & 56.78  & 31.85   & 44.35   & 1.65       \\
Villarreal          & 65.5         & 66.04  & 56.92   & 60.01   & 5.49       \\
Alav\'{e}s              & 46.1         & 52.45  & 31.11   & 42.83   & 3.27       \\
Getafe              & 44.5         & 52.58  & 32.63   & 43.25   & 1.25       \\
Eibar               & 46.3         & 55.30   & 31.99   & 44.87   & 1.43       \\
Legan\'{e}s             & 43.3         & 50.58  & 23.70    & 37.59   & 5.71      \\
                    &               &       &           &       &       
\end{tabular}

\textbf{Table 3}. La Liga 2017/2018. Actual allocation versus three alternative allocations.
\end{table}

In Table 3, 
La Liga's allocation for the 2017-2018 season is compared
with three allocations obtained by combining allocations from the four
dimensions, as follows. In Column 3 (Low SD), the
allocation with the lowest standard deviation for each dimension is selected. In Column 4 (High
SD), the allocation with the highest standard deviation for each dimension 
is selected. In Column 5 (Average), the average of the allocations
considered in such dimension is selected. For instance, in sport
performance, we compute the average between the three allocations considered
(the ones induced by La Liga, the Premier League, and the points). We
proceed in the same way for the other three dimensions. For each of the
three columns, the complete allocation is the sum over the allocations in
each dimension.

As the allocations with the lowest SD are more similar to the average, we
should expect that big clubs obtain more with the allocation High SD whereas
small clubs obtain more with Low SD. The first four clubs of the list obtain
more with High SD whereas the rest obtain more with Low SD.
One can also see that no club obtains more with the average than with the
maximum between Low SD and High SD. Nevertheless, Athletic Bilbao obtains
more with La Liga than with the maximum between Low SD and High SD. This
fact is remarkable because Low SD and High SD are some kind of extreme
allocations.

In column 6 of Table 3 (Difference) the difference between La Liga and the
Average column is computed. The club more favored by La Liga is Atl\'{e}tico
de Madrid obtaining $8.58$ more with La Liga. The worst treated club (by
far) is Betis, receiving $18.44$ less.
Betis is a widely popular club nationwide with large, which is reflected in the number of viewers its games have. 
During the 2017-18 season, Betis was mostly broadcasted in a free-to-air channel (as opposed to the ``Big 3", which were always broadcasted in subscription channels). We can only conjecture that La Liga rewarded more the audiences coming from games broadcasted in subscription channels, than those coming from free-to-air channels, to explain the disadvantage.

We conclude stressing that 
it would be interesting to perform similar analyses for other
important football leagues in Europe (such as the dominant English Premier
League), as well as 
for some professional leagues in the US, where the broadcasting rights are divided,
basically, via equal sharing (e.g., Fort and Quirk 1995).

\section{Conclusion}

Sports account for a large portion of all broadcasting attention. Although
times are changing and new generations shift towards other forms of
entertainment, sportscasting continues to be a major aspect of the
entertainment industry. Massive amounts of people worldwide consume sports
via broadcasting. And payments for broadcasting live sports
competitions have grown drastically over recent years, to the extent that
they have been shaping the role of sport rights in the broadcast industry
for a long time (e.g., Cave and Crandall, 2001).

We have reviewed in this survey the literature on the economics of
sportscast revenue sharing. We have concentrated on the focal case of
professional sports leagues (with a double round-robin format as the
benchmark case, although others could also be accommodated) in which
revenues are raised collectively. We have reviewed the literature on the
several approaches that exist to this problem; namely, a statistical
estimation approach, the axiomatic approach, an indirect approach to solve
the problems via associating a cooperative game and a decentralized approach
via majority voting. We have also explored extensions (via the operational
approach) to more general cases (for instance, those arising after leagues
are cancelled) and we have illustrated the analyses to the special case of
the Spanish Football League. We have also listed some open questions for
further research that arise in some of those approaches. Altogether, we can
safely argue that the economics of sharing the revenues from sportscasting
is a lively research topic nowadays.

This survey is mainly based on 11 recent papers we published. Table 4
summarizes the main contributions and results of each of these papers.

\begin{equation*}
\begin{tabular}{|l|l|}
\hline
\textbf{Paper} & \textbf{Summary} \\ \hline
B\&MT(2020a) & 
\begin{tabular}{l}
Introduction of the model. Axiomatic characterizations of the equal-split rule and concede-and-divide. \\ 
Alternative approaches: cooperative games, claim problems, statistical estimation. \\ 
Analysis of the fan effect. Application to La Liga.%
\end{tabular}
\\ \hline
B\&MT(2020b) & 
\begin{tabular}{l}
Introduction of axioms formalizing the notion of marginalism. 
\\ 
Axiomatic characterizations of several rules without resorting to additivity.%
\end{tabular}%
. \\ \hline
B\&MT(2021) & 
\begin{tabular}{l}
Axiomatic characterization of the family of compromise rules. Application to La Liga.
\\ 
Existence of a majority voting equilibrium for such a family and Lorenz comparisons.%
\end{tabular}
\\ \hline
B\&MT(2022a) & 
\begin{tabular}{l}
Comprehensive axiomatic study to uncover further the space of rules. 
\\ 
Axiomatic characterization of the family of generalized compromise rules.%
\end{tabular}
\\ \hline
B\&MT(2022b) & 
\begin{tabular}{l}
Introduction of axioms formalizing various forms of the principle of monotonicity. \\ 
Axiomatic characterizations of several rules based on those axioms.%
\end{tabular}
\\ \hline
B\&MT(2022c) & 
\begin{tabular}{l}
Axiomatic characterization of the rules satisfying additivity and a weak form of symmetry. \\ 
Axiomatic characterizations adding monotonicity axioms.%
\end{tabular}
\\ \hline
B\&MT(2023a) & 
\begin{tabular}{l}
Axiomatic characterization of the family of rules satisfying additivity and anonymity. \\ 
Majority voting equilibrium does not exist for that family but it does exist for a subfamily.%
\end{tabular}
\\ \hline
B\&MT(2023b) & 
\begin{tabular}{l}
Comprehensive application of our theoretical results to La Liga.\\ 
Alternatives in each of the four dimensions used by La Liga.%
\end{tabular}
\\ \hline
B\&MT(2023c) & 
\begin{tabular}{l}
Introduction of an extended model to analyze leagues with cancelled games via two methods. \\
Analysis of the extended equal-split rule and concede-and-divide via both methods.%
\end{tabular}
\\ \hline
B\&MT(2024a) & 
\begin{tabular}{l}
Comprehensive study of the principle of anonymity, compared with other impartiality axioms.
\\ 
Axiomatic characterizations of new rules based on anonymity.%
\end{tabular}
\\ \hline
B\&MT(2024b) & 
\begin{tabular}{l}
Analysis of the cooperative game associated with a broadcasting problem. \\ 
Axiomatic characterizations of the core and the Shapley value.%
\end{tabular}
\\ \hline
\end{tabular}
\end{equation*}

\textbf{Table 4}. Summary of results.
%

To conclude, we should argue that traditional television broadcasting is
gradually being replaced by streaming platforms. The rise of streaming
services and the emergence of the direct-to-consumer sports consumption
model has transformed the way people watch sports. For instance, the NBA
replaced long-term partner TNT with Amazon Prime as the third pole alongside
NBC and ESPN in its current broadcasting agreement. Amazon Prime already
owned Thursday Night Football. MLS and F1 recently turned to Apple, which
offered more money than any of its competitors. Peacock broadcasts several
sports, including College Football and the English Premier League. The list
goes on and sports leagues might see opportunities in the shifting
landscape, which might also affect the way in which revenues are shared
(among participating clubs). There exists by now an emerging literature
dealing with revenue sharing in streaming platforms (e.g., Alaei et al.,
2022; Lei, 2023; Berganti\~{n}os and Moreno-Ternero, 2025a, 2025b, 2026; Gon%
\c{c}alves-Dosantos et al., 2025a, 2025b) and it might be interesting to
explore the connections with this literature to account for eventual
economic aspects of sportscast revenue sharing that might arise in the near
future.


\end{document}